\documentclass[twocolumn,showpacs,preprintnumbers,amsmath,amssymb]{revtex4}
\usepackage{graphicx}
\usepackage{dcolumn}
\usepackage{bm}

\begin{document}

\preprint{APS/123-QED}

\title{A theoretical approach to thermal noise caused 
by an inhomogeneously distributed loss\\
--- Physical insight by the advanced modal expansion}

\author{Kazuhiro Yamamoto}
 \email{yamak@icrr.u-tokyo.ac.jp}
 \altaffiliation[Present address: ]{Gravitational wave group, 
Institute for Cosmic Ray Research, 
the University of Tokyo,
5-1-5 Kashiwa-no-Ha, Kashiwa, Chiba 277-8582, Japan.}
\author{Masaki Ando}
\author{Keita Kawabe}
\author{Kimio Tsubono}
\affiliation{%
Department of Physics, the University of Tokyo, 
7-3-1 Hongo, Bunkyo-ku, Tokyo 113-0033, Japan}%


\date{\today}

\begin{abstract} 
We modified the modal expansion, which is the traditional method 
used to calculate thermal noise. 
This advanced modal expansion provides 
physical insight about the discrepancy 
between the actual thermal noise caused 
by inhomogeneously distributed loss and the 
traditional modal expansion. 
This discrepancy comes from correlations between the 
thermal fluctuations of the resonant modes. 
The thermal noise spectra estimated by 
the advanced modal expansion are consistent 
with the results of measurements of 
thermal fluctuations caused by inhomogeneous losses. 
\end{abstract}


\maketitle

\section{Introduction}

Thermal fluctuation is one of the fundamental noise sources 
in precise measurements. 
For example, the sensitivity of interferometric gravitational wave detectors 
\cite{LIGO,VIRGO,GEO,TAMA} is limited by the thermal noise of 
the mechanical components. 
The calculated thermal fluctuations of rigid cavities 
have coincided with the highest laser 
frequency stabilization results ever obtained \cite{Numata5,Notcutt}.  
It is important to evaluate the thermal motion 
for studying the noise property. 
The (traditional) modal expansion \cite{Saulson} has been commonly 
used to calculate the thermal noise of elastic systems. 
However, recent experiments 
\cite{Yamamoto1,Harry,Conti,Numata3,Yamamoto3,Black} have 
revealed that modal expansion is not correct 
when the mechanical dissipation is distributed inhomogeneously. 
In some theoretical studies \cite{Levin,Nakagawa1,Tsubono,Yamamoto-D}, 
calculation methods that are completely different 
from modal expansion have been developed. These methods are supported 
by the experimental results of 
inhomogeneous loss \cite{Harry,Numata3,Yamamoto3,Black}. 
However, even when these method were used, 
the physics of the discrepancy between 
the actual thermal noise and the traditional modal expansion 
was not fully understood. 

In this paper, 
another method to calculate the thermal noise 
is introduced \cite{Yamamoto-D}. 
This method, advanced modal expansion, is a modification of 
the traditional modal expansion
(this improvement is a general extension of 
a discussion in Ref. \cite{Majorana}). 
The thermal noise spectra estimated by 
this method are consistent with the results of 
experiments concerning inhomogeneous 
loss \cite{Yamamoto1, Yamamoto3}. 
It provides information about the disagreement 
between the thermal noise and the traditional modal expansion. 
We present the details of these topics in the following sections.   

\section{Outline of advanced modal expansion}

\subsection{Review of the traditional modal expansion}

The thermal fluctuation of 
the observed coordinate, $X$, of a linear 
mechanical system is derived 
from the fluctuation-dissipation theorem \cite{Callen,Greene,Landau2}, 
\begin{eqnarray}
G_{X}(f)&=&-\frac{4 k_{\rm B} T}{\omega} {\rm Im}[H_{X}(\omega)],
\label{FDT}\\
H_{X}(\omega)&=&\frac{\tilde{X}(\omega)}{\tilde{F}(\omega)},
\label{transfer function}\\
\tilde{X}(\omega)&=&\frac1{2\pi}
\int^{\infty}_{-\infty}X(t)\exp(-{\rm i}\omega t)dt,
\label{Fourier transform}
\end{eqnarray}
where $f(=\omega/2\pi)$, $t$, $k_{\rm B}$ and $T$, are the frequency, time, 
Boltzmann constant and temperature, respectively. 
The functions ($G_{X}$, $H_{X}$, and $F$) are the (single-sided) 
power spectrum density of 
the thermal fluctuation of $X$, the transfer function, 
and the generalized force, which corresponds to $X$. 
In the traditional modal expansion \cite{Saulson}, 
in order to evaluate this transfer function, 
the equation of motion of the mechanical system 
without any loss is decomposed into those of the resonant modes. 
The details are as follows: 

\begin{figure}
\includegraphics[width=8.6cm]{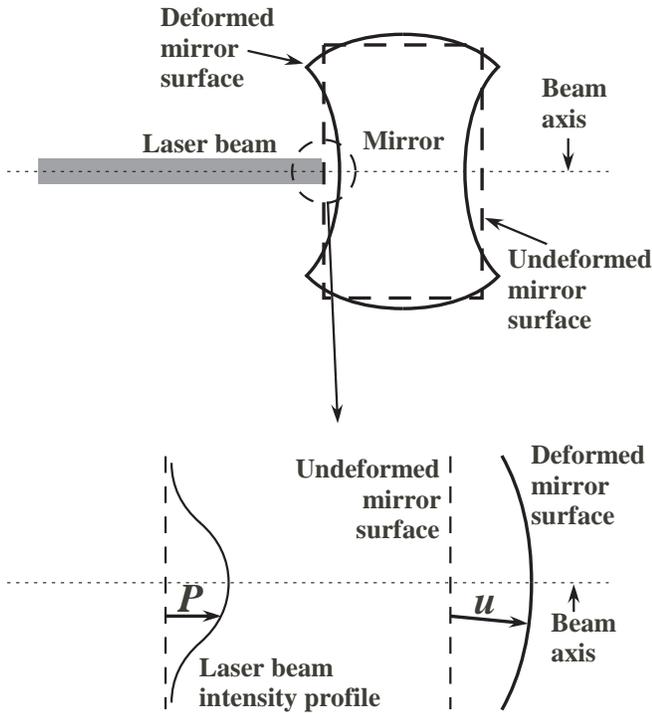}
\caption{\label{defX}Example of the definition 
of the observed coordinate, $X$, in Eq. (\ref{observed coordinate}). 
The mirror motion is observed using a Michelson interferometer. 
The coordinate $X$ is the output of the interferometer. 
The vector $\boldsymbol{u}$ represents the displacement 
of the mirror surface.
The field $\boldsymbol{P}$ is parallel to the beam axis.
Its norm is the beam-intensity profile \cite{Levin}.}
\end{figure}
The definition of the observed coordinate, $X$, is described as 
\begin{equation}
X(t) = \int \boldsymbol{u}(\boldsymbol{r},t) 
\cdot \boldsymbol{P}(\boldsymbol{r}) dS,
\label{observed coordinate}
\end{equation}
where $\boldsymbol{u}$ is the displacement of the system and 
$\boldsymbol{P}$ is a
weighting function that describes 
where the displacement is measured.
For example, 
when mirror motion is observed using a Michelson
interferometer, as in Fig. \ref{defX}, $X$ and 
$\boldsymbol{u}$ represent the interferometer output and 
the displacement of the mirror surface, respectively.
The vector $\boldsymbol{P}$ is parallel to the beam axis.
Its norm is the beam-intensity profile \cite{Levin}. 
The equation of motion of the mechanical system without dissipation 
is expressed as
\begin{equation}
\rho\frac{\partial^2 \boldsymbol{u}}{\partial t^2}
-{\cal L}[\boldsymbol{u}]=F(t)\boldsymbol{P}(\boldsymbol{r}),
\label{eq_mo_continuous}
\end{equation}  
where $\rho$ is the density and ${\cal L}$ is a linear operator.
The first and second terms 
on the left-hand side of Eq. (\ref{eq_mo_continuous}) 
represent the inertia and the restoring force 
of the small elements in the mechanical oscillator, respectively.  
The solution of Eq. (\ref{eq_mo_continuous}) is the 
superposition of the basis functions,  
\begin{equation}
\boldsymbol{u}(\boldsymbol{r},t)
=\sum_{n}\boldsymbol{w}_n(\boldsymbol{r})q_n(t).
\label{mode decomposition}
\end{equation}
The functions, $\boldsymbol{w}_n$ and $q_n$, 
represent the displacement and time development of the $n$-th 
resonant mode, respectively.
The basis functions, $\boldsymbol{w}_n$, 
are solutions of the eigenvalue problem, written as
\begin{equation}
{\cal L}[\boldsymbol{w}_n(\boldsymbol{r})]
=-\rho {\omega_n}^2 \boldsymbol{w}_n(\boldsymbol{r}),
\label{eigenvalue problem}
\end{equation}
where $\omega_n$ is 
the angular resonant frequency of the 
$n$-th mode.
The displacement, $\boldsymbol{w}_n$, 
is the component of an orthogonal complete system, 
and is normalized to satisfy the condition
\begin{equation}
\int \boldsymbol{w}_n(\boldsymbol{r}) 
\cdot \boldsymbol{P}(\boldsymbol{r}) dS = 1.
\label{normalized condition}
\end{equation}
The formula of the orthonormality is described as
\begin{equation}
\int \rho \boldsymbol{w}_n(\boldsymbol{r}) \cdot 
\boldsymbol{w}_k(\boldsymbol{r}) dV
= m_n \delta_{nk}.
\label{effective mass} 
\end{equation}
The parameter  
$m_n$ is called the effective mass of the mode 
\cite{Yamamoto1, Gillespie, Bondu, Yamamoto2}. 
The tensor $\delta_{nk}$ is the Kronecker's $\delta$-symbol.

Putting Eq. (\ref{mode decomposition}) 
into Eq. (\ref{observed coordinate}),  
we obtain a relationship between $X$ and $q_n$ 
using Eq. (\ref{normalized condition}),
\begin{equation}
X(t) = \sum_{n} q_n(t).
\label{observed coordinate decomposition}
\end{equation} 
In short, coordinate $X$ 
is a superposition of those of the modes, $q_n$. 
In order to decompose the equation of motion, 
Eq. (\ref{eq_mo_continuous}), 
Eq. (\ref{mode decomposition}) is substituted for $\boldsymbol{u}$ in 
Eq. (\ref{eq_mo_continuous}).
Equation (\ref{eq_mo_continuous}) is multiplied by $\boldsymbol{w}_n$ 
and then integrated over all of the volume using 
Eqs. (\ref{eigenvalue problem}),  
(\ref{normalized condition}) and (\ref{effective mass}). 
The result is that the equation of motion of the $n$-th mode, $q_n$, 
is the same as
that of a harmonic oscillator on which force $F(t)$ is applied.
After modal decomposition, 
the dissipation term is added to the equation of each mode.
The equation of the $n$-th mode is written as 
\begin{equation}
-m_n \omega^2 \tilde{q}_n + m_n {\omega_n}^2 [1+{\rm i}\phi_n(\omega)] 
\tilde{q}_n=\tilde{F},
\label{traditional1}
\end{equation}
in the frequency domain. 
The function $\phi_n$ is the loss angle, which represents 
the dissipation of the $n$-th mode \cite{Saulson}. 
The transfer function, $H_X$, derived 
from Eqs. (\ref{transfer function}), 
(\ref{observed coordinate decomposition}) and 
(\ref{traditional1}) is the summation of those of the modes, $H_n$, 
\begin{eqnarray}
H_{X}(\omega) &=& \frac{\tilde{X}}{\tilde{F}} 
= \sum_n \frac{\tilde{q}_n}{\tilde{F}} \left(= \sum_n H_n \right)\nonumber\\
&=& \sum_{n} \frac1{-m_n \omega^2  
+ m_n {\omega_n}^2 [1+{\rm i}\phi_n(\omega)]}.
\label{traditional3}
\end{eqnarray}
According to Eqs. (\ref{FDT}) and (\ref{traditional3}), 
the power spectrum density, $G_{X}$, is 
the summation of the power spectrum, $G_{q_n}$, of $q_n$, 
\begin{eqnarray}
&&G_{X}(f) = \sum_{n} G_{q_n}\nonumber\\
&&= \sum_{n} \frac{4k_{\rm B}T}{m_n \omega}
\frac{{\omega_n}^2\phi_n(\omega)}
{(\omega^2-{\omega_n}^2)^2+{\omega_n}^4{\phi_n}^2(\omega)}.
\label{traditional2} 
\end{eqnarray}      

\subsection{Equation of motion in an advanced modal expansion}

In the traditional modal expansion, the dissipation term is introduced 
after decomposition of the equation of motion without any loss. 
On the contrary, in an advanced modal expansion, the equation 
with the loss is decomposed \cite{Yamamoto-D,optics}. 
If the loss is sufficiently small, the expansion process is similar to 
that in the perturbation theory of quantum mechanics 
\cite{Sakurai}. 
The equation of $q_n$ is expressed as 
\begin{eqnarray}
-m_n \omega^2 \tilde{q}_n + m_n {\omega_n}^2 [1&+&{\rm i}\phi_n(\omega)] 
\tilde{q}_n\nonumber\\
&+& \sum_{k \neq n} {\rm i} \alpha_{nk}(\omega) \tilde{q}_k 
= \tilde{F},\label{advanced1}\\
\phi_n(\omega) &=& \frac{\alpha_{nn}}{m_n {\omega_n}^2}\label{phi}.
\label{phi_n}
\end{eqnarray}
The third term in Eq. (\ref{advanced1}) is the difference 
between the advanced, Eq. (\ref{advanced1}), 
and traditional, Eq. (\ref{traditional1}), modal expansions. 
Since this term is a linear combination of the motions of the other modes, 
it represents the couplings between the modes. 
The magnitude of the coupling, $\alpha_{nk}$, depends on the property and 
the distribution of the loss (described below). 

\subsection{Details of coupling}

Let us consider the formulae of the couplings caused 
by the typical inhomogeneous losses, the origins of which exist 
outside and inside  
the material (viscous damping and structure damping, respectively) 
\cite{Yamamoto-D}.
Regarding most of the external losses, 
for example, the eddy-current damping and residual gas damping are of 
the viscous type \cite{Saulson}. The friction force of this damping is 
proportional to the velocity. 
Inhomogeneous viscous damping introduces a friction force, 
${\rm i} \omega \rho \Gamma(\boldsymbol{r}) 
\tilde{\boldsymbol{u}}(\boldsymbol{r})$,
into the left-hand side of 
the equation of motion, Eq. (\ref{eq_mo_continuous}), 
in the frequency domain. The function 
$\Gamma (\geq 0)$ represents the strength of the damping. 
The equation of motion with the dissipation term, 
${\rm i} \omega \rho \Gamma(\boldsymbol{r}) 
\tilde{\boldsymbol{u}}(\boldsymbol{r})$, 
is decomposed. Since the loss is small, the basis functions 
of the equation without loss 
are available \cite{Sakurai}.
Equation (\ref{mode decomposition}) is put into 
the equation of motion along with the inhomogeneous viscous damping.
This equation multiplied by $\boldsymbol{w}_n$
is integrated.
The coupling of this dissipation is written in the form 
\begin{equation}
\alpha_{nk} = \omega \int \rho \Gamma(\boldsymbol{r}) 
\boldsymbol{w}_n(\boldsymbol{r}) \cdot  
\boldsymbol{w}_k(\boldsymbol{r})  dV = \alpha_{kn}.
\label{coupling_viscous}  
\end{equation}

In most cases, the internal loss in the material is expressed 
using the phase lag, $\phi (\geq 0)$, 
between the strain and the stress \cite{Saulson}.
The magnitude of the dissipation is proportional to this lag. 
The phase lag is almost constant against the frequency \cite{Saulson} 
in many kinds of materials (structure damping).
In the frequency domain, the relationship between the strain and the stress 
(the generalized Hooke's law)
in an isotropic elastic body 
is written as \cite{Saulson,Levin,Yamamoto-D,Landau} 
\begin{eqnarray}
\tilde{\sigma}_{ij} 
&=& \frac{E_0[1+{\rm i}\phi(\boldsymbol{r})]}{1+\sigma}
\left(\tilde{u}_{ij} + \frac{\sigma}{1-2\sigma}\sum_{l}\tilde{u}_{ll}
\delta_{ij}\right)\nonumber\\
&=& [1+{\rm i}\phi(\boldsymbol{r})]\tilde{\sigma}'_{ij}, 
\label{structure_stress}\\
u_{ij} &=& \frac1{2} \left(\frac{\partial u_i}{\partial x_j} 
+ \frac{\partial u_j}{\partial x_i}\right),
\label{strain}
\end{eqnarray}
where $E_0$ is Young's modulus and $\sigma$ is the Poisson ratio; 
$\sigma_{ij}$ and $u_{ij}$ are the stress and strain tensors, respectively.
The tensor, $\tilde{\sigma}'_{ij}$, is the real part of 
the stress, $\tilde{\sigma}_{ij}$. It represents 
the stress when the structure damping vanishes. 
The value, $u_i$, is the $i$-th component of  
$\boldsymbol{u}$. 
The equation of motion of an elastic body 
\cite{Landau} in the frequency domain is expressed as  
\begin{equation}
-\rho \omega^2 \tilde{u}_i
-\sum_j \frac{\partial \tilde{\sigma}_{ij}}{\partial x_{j}}
= \tilde{F}P_i(\boldsymbol{r}), 
\label{eq_mo_elastic_withloss}
\end{equation} 
where $P_i$ is the $i$-th component of  
$\boldsymbol{P}$. 
From Eqs. (\ref{structure_stress}) and (\ref{eq_mo_elastic_withloss}), 
an inhomogeneous structure damping term is obtained, 
$-{\rm i} \sum_j \partial \phi(\boldsymbol{r}) 
\tilde{\sigma}'_{ij}/\partial x_{j}$.
The equation of motion with the inhomogeneous structure damping 
is decomposed in the same manner as that of the inhomogeneous 
viscous damping.
The coupling is calculated using integration by parts and Gauss' theorem 
\cite{Landau}, 
\begin{widetext}
\begin{eqnarray}
\alpha_{nk} &=& - \int \sum_{i,j} w_{n,i} 
\frac{\partial \phi(\boldsymbol{r}) \sigma_{k,ij}}{\partial x_j} dV
\nonumber\\
&=& - \int \sum_{i,j} 
\frac{\partial w_{n,i} \phi(\boldsymbol{r}) \sigma_{k,ij}}
{\partial x_j} dV
+ \int \sum_{i,j} \frac{\partial w_{n,i}}{\partial x_j} 
\phi(\boldsymbol{r}) \sigma_{k,ij} dV \nonumber \\
&=& - \int \sum_{i,j} w_{n,i} \phi(\boldsymbol{r}) \sigma_{k,ij} n_j dS
+ \int \sum_{i,j} \frac{\partial w_{n,i}}{\partial x_j} 
\phi(\boldsymbol{r}) \sigma_{k,ij} dV \nonumber \\
&=& \int \frac{E_0 \phi(\boldsymbol{r})}{1+\sigma} 
\left[\sum_{i,j} \frac{\partial w_{n,i}}{\partial x_j}
\left(w_{k,ij} + \frac{\sigma}{1-2\sigma} 
\sum_l w_{k,ll} \delta_{ij}\right)\right] dV\nonumber\\
&=& \int \frac{E_0 \phi(\boldsymbol{r})}{1+\sigma}
\left(\sum_{i,j}w_{n,ij}w_{k,ij}
+ \frac{\sigma}{1-2\sigma}\sum_{l}w_{n,ll}\sum_{l}w_{k,ll} \right) dV
=\alpha_{kn},
\label{coupling_structure}
\end{eqnarray}
\end{widetext}
where $w_{n,i}$ and $n_{i}$ are the $i$-th components 
of $\boldsymbol{w}_n$ and the normal unit vector on the surface. 
The tensors, $w_{n,ij}$ and $\sigma_{n,ij}$, are the strain and stress tensors 
of the $n$-th mode, respectively. In order to calculate these tensors, 
$w_{n,i}$ is substituted for $u_i$ 
in Eqs. (\ref{structure_stress}) and (\ref{strain}) with $\phi=0$.  
Equation (\ref{coupling_structure}) is valid 
when the integral of the function, 
$\sum_{i,j} w_{n,i} \phi \sigma_{k,ij} n_{j}$, 
on the surface of the elastic body vanishes. 
For example, the surface is fixed 
($w_{n,i}=0$) or free ($\sum_{j}\sigma_{k,ij}n_{j}=0$) \cite{Landau}.

The equation of motion in the advanced modal expansion coincides with 
that in the traditional modal expansion when all of the 
couplings vanish.
A comparison between 
Eqs. (\ref{effective mass}) and (\ref{coupling_viscous}) shows that 
in viscous damping all $\alpha_{nk} (n \neq k)$ are zero 
when the dissipation strength, $\Gamma(\boldsymbol{r})$, 
does not depend on the position, $\boldsymbol{r}$.  
In the case of structure damping, from Eqs. (\ref{eq_mo_continuous}), 
(\ref{mode decomposition}), 
(\ref{eigenvalue problem}) and (\ref{eq_mo_elastic_withloss}), 
the stress, $\sigma'_{ij}$, 
without dissipation satisfies 
\begin{equation}
\sum_j \frac{\partial \tilde{\sigma}'_{ij}}{\partial x_{j}}
= - \sum_n \rho {\omega_n}^2 w_{n,i} \tilde{q}_n.
\label{stress decomposition}
\end{equation} 
According to Eq. (\ref{effective mass}), Eq. (\ref{stress decomposition}) 
is decomposed without any couplings.
From Eq. (\ref{stress decomposition}) 
and the structure damping term, $-{\rm i} \sum_j \partial \phi(\boldsymbol{r}) 
\tilde{\sigma}'_{ij}/\partial x_{j}$, 
the conclusion is derived; all of the couplings in the structure damping 
vanish when the loss amplitude, $\phi$, is homogeneous. 
In summary, the inhomogeneous viscous and structure dampings produce 
mode couplings
and destroy the traditional modal expansion.

The reason why the inhomogeneity of the loss causes the couplings 
is as follows. Let us consider the decay motion after only one 
resonant mode 
is excited. If the loss is uniform, 
the shape of the displacement of the system 
does not change while the resonant motion decays. 
On the other hand, 
if the dissipation is inhomogeneous, the motion near 
the concentrated loss decays more rapidly than the other parts.   
The shape of the displacement becomes different from that of 
the original resonant mode. 
This implies that the other modes are excited, i.e. 
the energy of the original mode is leaked to the other modes. 
This energy leakage represents the couplings in the equation of motion.

It must be noticed that some kinds of 
"homogeneous" loss cause the couplings. 
For example, in thermoelastic damping 
\cite{Zener,Braginsky-thermo,Liu,Cerdonio}, 
which is a kind of internal loss, 
the energy components of the shear strains, $w_{n,ij}(i \neq j)$, 
are not dissipated. The couplings, $\alpha_{nk}$, 
do not have any terms that consist of the shear strain tensors.
The coupling formula of the homogeneous thermoelastic damping is different 
from Eq. (\ref{coupling_structure}) with the constant $\phi$. 
The couplings are not generally zero, even if the thermoelastic damping is 
uniform. The advanced, not traditional, modal expansion provides a 
correct evaluation of the "homogeneous" thermoelastic damping. 
In this paper, however, only coupling caused by inhomogeneous loss 
is discussed. 

\subsection{Thermal-noise formula of advanced modal expansion}
\label{thermal noise of advanced}

In the advanced modal expansion, the transfer function, $H_{X}$, 
is derived from Eqs. (\ref{transfer function}), 
(\ref{observed coordinate decomposition}), 
and (\ref{advanced1})
(since the dissipation is small, 
only the first-order of $\alpha_{nk}$ is considered \cite{alpha2}),
\begin{widetext}
\begin{equation}
H_{X}(\omega) = \sum_n \frac1{-m_n {\omega}^2 
+ m_n {\omega_n}^2 (1 + {\rm i}\phi_n)}
- \sum_{k \neq n} 
\frac{{\rm i} \alpha_{nk}}
{[-m_n \omega^2+m_n {\omega_n}^2 (1+{\rm i}\phi_n)]
[-m_k \omega^2 +m_k {\omega_k}^2 (1+{\rm i}\phi_k)]}.
\label{advanced3}
\end{equation}
\end{widetext}
Putting Eq. (\ref{advanced3}) 
into Eq. (\ref{FDT}), 
the formula for the thermal noise is obtained. 
In the off-resonance region, where 
$|-\omega^2+{\omega_n}^2| \gg {\omega_n}^2 \phi_n(\omega)$ for all $n$, 
this formula approximates the expression 
\begin{eqnarray}
G_{X}(f)&=&\sum_{n} \frac{4 k_{\rm B} T}{m_n\omega}
\frac{{\omega_n}^2\phi_n(\omega)}
{(\omega^2-{\omega_n}^2)^2}\nonumber\\
&+&\sum_{k \neq n}\frac{4k_{\rm B}T}{m_n m_k \omega}
\frac{\alpha_{nk}}{(\omega^2-{\omega_n}^2)(\omega^2-{\omega_k}^2)}.
\label{advanced2}
\end{eqnarray}
The first term is the same as the formula of 
the traditional modal expansion, 
Eq. (\ref{traditional2}). 

The interpretation of Eq. (\ref{advanced2}) is as follows. 
The power spectrum density of the thermal fluctuation force 
of the $n$-th mode, $G_{F_n}$, 
and the cross-spectrum density between $F_n$ and $F_k$, $G_{F_n F_k}$, are 
evaluated from Eq. (\ref{advanced1}) and the fluctuation-dissipation theorem
\cite{Greene,Landau2},
\begin{eqnarray}
G_{F_n}(f) &=& 4 k_{\rm B} T 
\frac{m_n {\omega_n}^2 \phi_n(\omega)}{\omega},
\label{G_F_n}\\
G_{F_n F_k}(f) &=& 4 k_{\rm B} T \frac{\alpha_{nk}(\omega)}{\omega}.
\label{G_F_n_F_k}
\end{eqnarray}
The power spectrum density, $G_{F_n}$, is independent of $\alpha_{nk}$. 
On the other hand, 
$G_{F_n F_k}$ depends on $\alpha_{nk}$. 
Having the correlations between the fluctuation forces 
of the modes, correlations between the motion of the
modes must also exist. 
The power spectrum density of the 
fluctuation of $q_n$, $G_{q_n}$, 
and the cross-spectrum density 
between the fluctuations of $q_n$ and $q_k$, 
$G_{q_nq_k}$, are described as \cite{Greene,Landau2}
\begin{eqnarray}
G_{q_n}(f) &=& \frac{4 k_{\rm B} T}{m_n\omega}\frac{{\omega_n}^2\phi_n(\omega)}
{(\omega^2-{\omega_n}^2)^2},
\label{G_q_n}\\
G_{q_n q_k}(f) &=& \frac{4k_{\rm B}T}{m_nm_k\omega}
\frac{\alpha_{nk}}{(\omega^2-{\omega_n}^2)(\omega^2-{\omega_k}^2)}, 
\label{G_q_n_q_k}
\end{eqnarray}    
under the same approximation of Eq. (\ref{advanced2}).
The first and second terms in Eq. (\ref{advanced2}) are summations of the 
fluctuation motion of each mode, Eq. (\ref{G_q_n}), 
and the correlations, Eq. (\ref{G_q_n_q_k}), respectively.  
In conclusion, inhomogeneous mechanical dissipation causes  
mode couplings and correlations of the thermal motion between the modes. 

In order to check wheather the formula of the thermal motion 
in the advanced modal expansion 
is consistent with the the equipartition principle, 
the mean square 
of the thermal fluctuation, $\overline{X^2}$, which 
is an integral of the power spectrum density 
over the whole frequency region, is evaluated. 
This mean square is derived from Eq. (\ref{FDT}) 
using the Kramers-Kronig relation \cite{Landau2,KKcomment},
\begin{equation}
{\rm Re}[H_X(\omega)] = -\frac1{\pi} \int_{-\infty}^{\infty} 
\frac{{\rm Im}[H_X(\xi)]}{\xi-\omega}d\xi.
\label{Kramers-Kronig}
\end{equation}
The calculation used to evaluate the mean square is written as \cite{Landau2} 
\begin{eqnarray}
\overline{X^2} &=& \int_0^{\infty} G_{X}(f) df \nonumber\\
&=& \frac1{4\pi} \int_{-\infty}^{\infty} G_{X}(\omega) d\omega \nonumber\\
&=& -\frac{k_{\rm B}T}{\pi} \int_{-\infty}^{\infty} 
\frac{{\rm Im}[H_{X}(\omega)]}{\omega} d\omega \nonumber\\
&=& k_{\rm B}T {\rm Re}[H_{X}(0)].
\label{X2}
\end{eqnarray}
Since the transfer function, $H_{X}$, is the ratio of 
the Fourier components of the real functions, 
the value $H_X(0)$ is a real number. The functions 
$\phi_n$ and $\alpha_{nk}$, which cause the imaginary 
part of $H_{X}$, must vanish when $\omega$ is zero \cite{Landau2}.  
The correlations do not affect the mean square of the 
thermal fluctuation. Equation (\ref{X2}) is rewritten using 
Eq. (\ref{advanced3}) as 
\begin{equation}
\overline{X^2}=\sum_n \frac{k_{\rm B}T}{m_n {\omega_n}^2}.
\label{X2_2}
\end{equation}
Equation (\ref{X2_2}) is equivalent to the prediction of 
the equipartition principle. 

The calculation of the formula of the advanced modal expansion, 
Eq. (\ref{advanced2}), is more troublesome than that of the other 
methods \cite{Levin,Nakagawa1,Tsubono,Yamamoto-D}, 
which are completely different from the modal expansion,  
when many modes contribute to the thermal motion.
However, the advanced modal expansion gives clear physical insight 
about the discrepancy between the thermal motion and the traditional modal 
expansion, as shown in Sec. \ref{new insight}. It is difficult to find 
this insight using other methods. 

\section{Experimental check}

\begin{figure*}
\begin{minipage}{8.6cm}
\begin{minipage}{3cm}
\includegraphics[width=3cm]{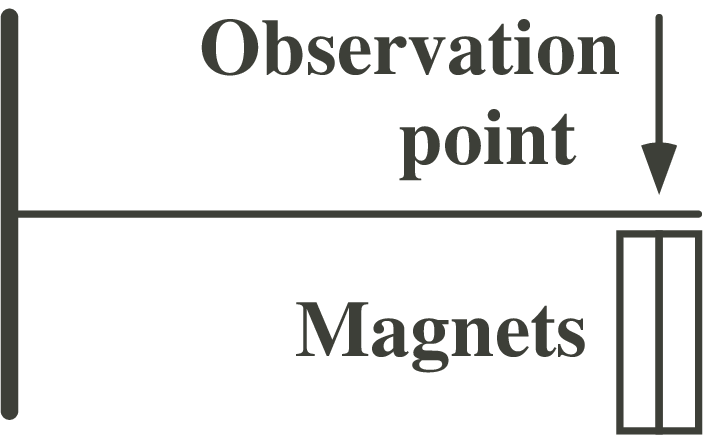}
\end{minipage}
\quad
\begin{minipage}{8.6cm}
\includegraphics[width=8.6cm]{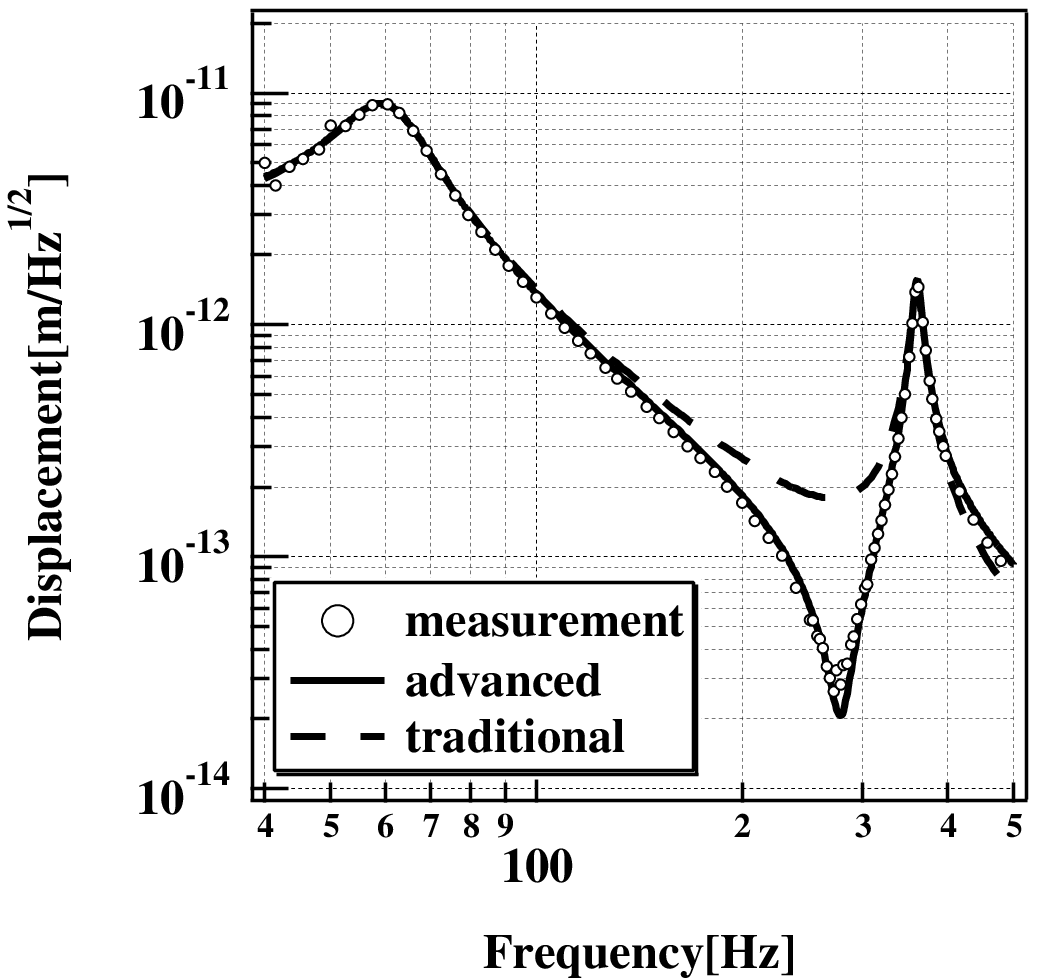}
\end{minipage}
\end{minipage}
\quad
\begin{minipage}{8.6cm}
\begin{minipage}{3cm}
\includegraphics[width=3cm]{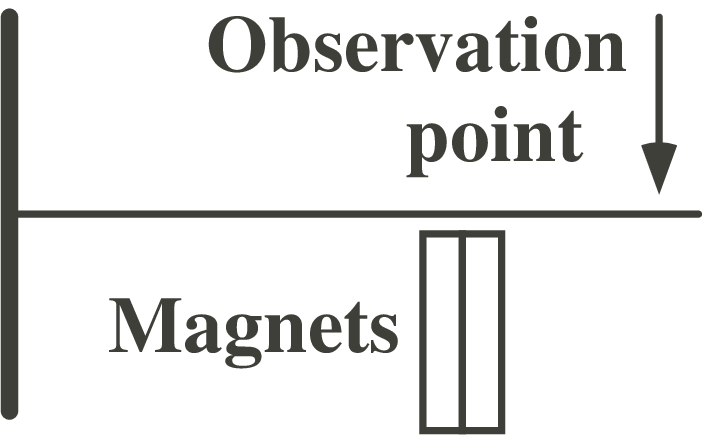}
\end{minipage}
\quad
\begin{minipage}{8.6cm}
\includegraphics[width=8.6cm]{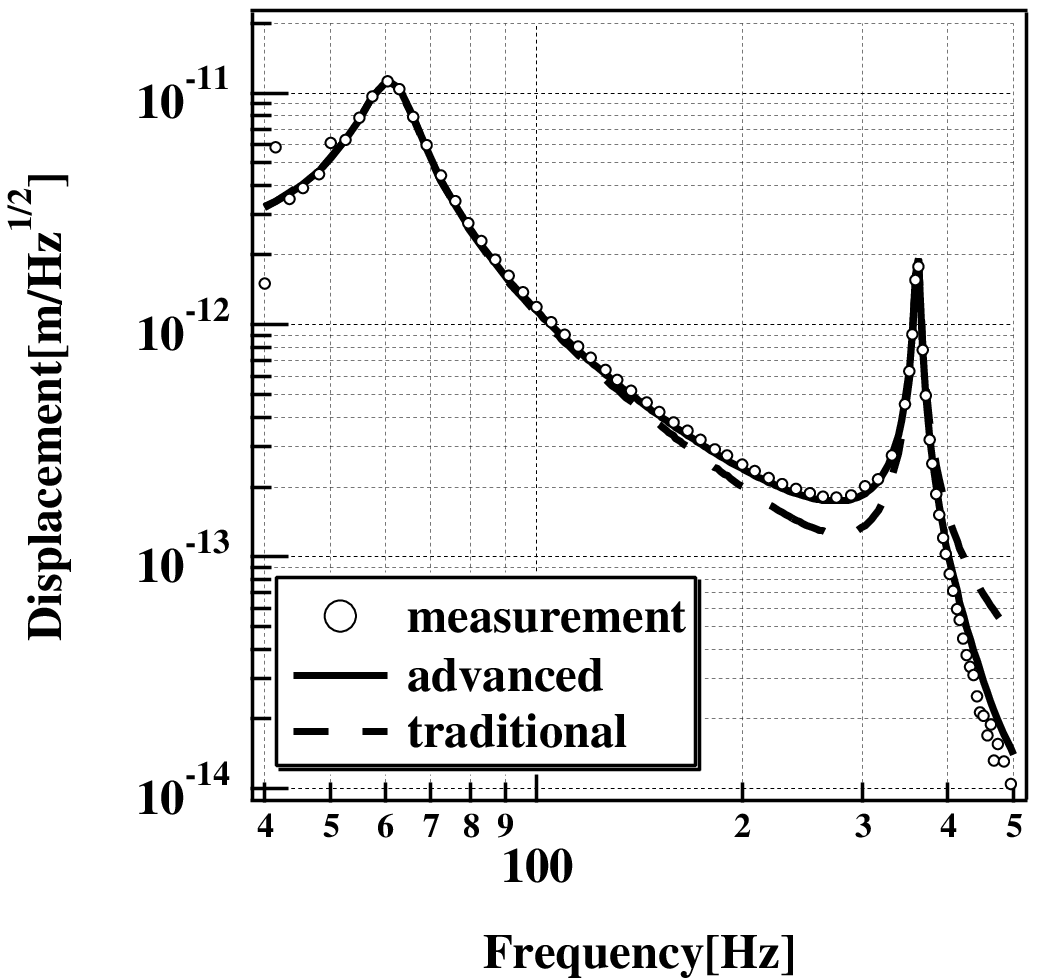}
\end{minipage}
\end{minipage}
\end{figure*}
\begin{figure*}
\begin{minipage}{8.6cm}
\begin{minipage}{3cm}
\includegraphics[width=3cm]{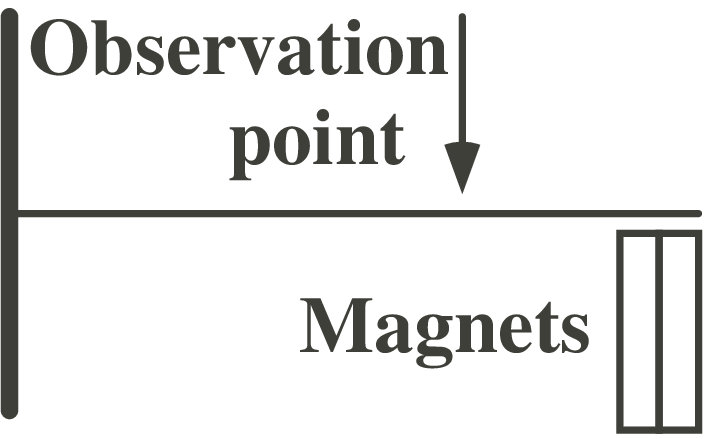}
\end{minipage}
\quad
\begin{minipage}{8.6cm}
\includegraphics[width=8.6cm]{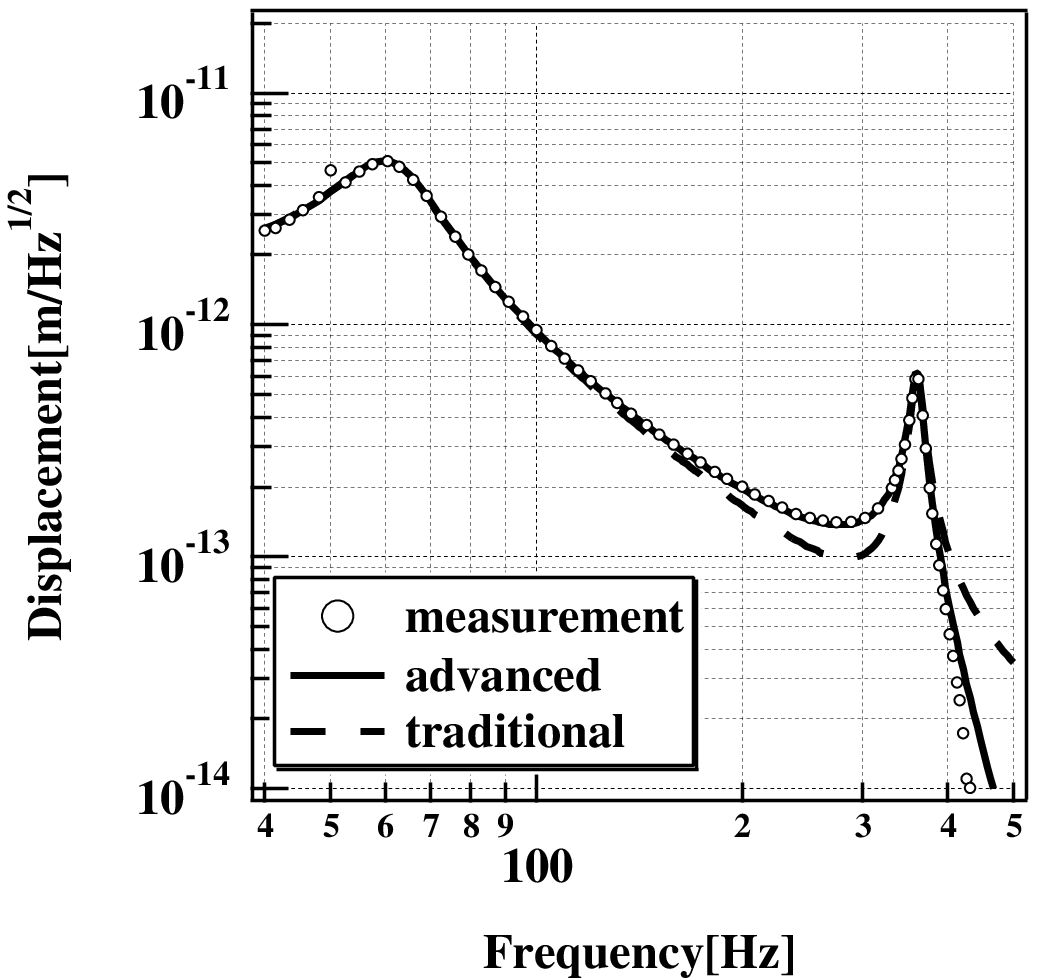}
\end{minipage}
\end{minipage}
\quad
\begin{minipage}{8.6cm}
\begin{minipage}{3cm}
\includegraphics[width=3cm]{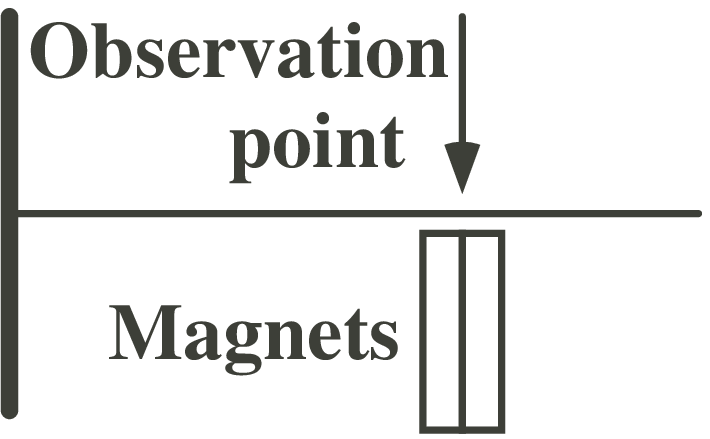}
\end{minipage}
\quad
\begin{minipage}{8.6cm}
\includegraphics[width=8.6cm]{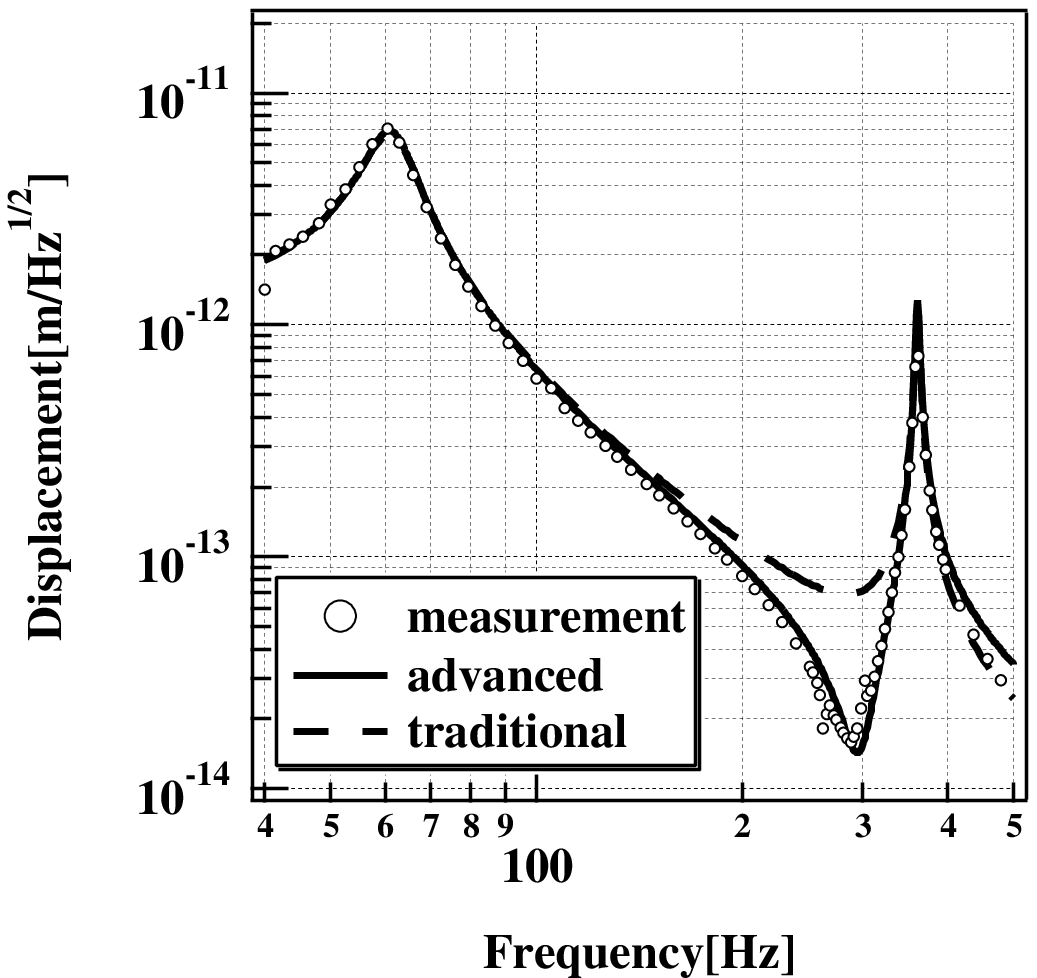}
\end{minipage}
\end{minipage}
\caption{\label{experiment}Comparison between the estimated 
advanced modal expansion and the experimental results of 
an aluminum alloy leaf spring 
with inhomogeneous eddy-current damping \cite{Yamamoto1}. 
The position of the 
magnets for the eddy-current damping and the observation 
point are indicated above each graph. 
In the figures above each graph, 
the left side of the leaf spring is fixed. 
The right side is free. 
The open circles in the graphs represent the power spectra 
of the thermal motion 
derived from the measured transfer functions using the fluctuation-dissipation 
theorem. These values coincide with the directly measured  
thermal-motion spectra \cite{Yamamoto1}. 
The solid lines are estimations using 
the advanced modal expansion. 
As a reference, an evaluation of the 
traditional modal expansion is also given (dashed lines).}
\end{figure*}
In order to test the advanced modal expansion experimentally, 
our previous experimental results concerning oscillators with 
inhomogeneous losses \cite{Yamamoto1,Yamamoto3} are 
compared with an evaluation of 
the advanced modal expansion \cite{Yamamoto-D}. 
In an experiment involving a drum 
(a hollow cylinder made from aluminum alloy 
as the prototype of the mirror in the interferometer) 
with inhomogeneous eddy-current damping by magnets 
\cite{Yamamoto3}, 
the measured values agreed
with the formula of the direct approach \cite{Levin}, 
Eq. (6) in Ref. \cite{Yamamoto3}. This expression is the same as that 
of the advanced modal expansion \cite{Yamamoto-D}. 

Figure \ref{experiment} presents the measured spectra 
of an aluminum alloy leaf spring 
with inhomogeneous eddy-current damping \cite{Yamamoto1}. 
The position of the 
magnets for the eddy-current damping and the observation 
point are indicated above each graph. 
In the figures above each graph, 
the left side of the leaf spring is fixed. 
The right side is free. 
The open circles in the graphs represent the power spectra of 
the thermal motion 
derived from the measured transfer functions using the fluctuation-dissipation 
theorem. These values coincide with the directly measured  
thermal-motion spectra \cite{Yamamoto1}. 
The solid lines are estimations using 
the advanced modal expansion 
(the correlations derived from Eqs. (\ref{coupling_viscous}) and 
(\ref{G_q_n_q_k})
are almost perfect \cite{Yamamoto-D}). 
As a reference, an evaluation of the 
traditional modal expansion is also given (dashed lines). 
The results of a leaf-spring experiment are consistent 
with the advanced modal expansion. 
Therefore, our two experiments support the advanced modal expansion.

\section{Physical insight given by the advanced modal expansion}
\label{new insight}

The advanced modal expansion provides physical insight about the 
disagreement between the real thermal motion 
and the traditional modal expansion. 
Here, let us discuss the three factors 
that affect this discrepancy:  
the number of the modes, the absolute value and the sign of the correlation. 

\subsection{Number of modes}

Since the difference between the advanced 
and traditional modal expansions is 
the correlations between the multiple modes, 
the number of the modes affects the magnitude of the discrepancy. 
If the thermal fluctuation is dominated by the contribution of only one mode, 
this difference is negligible, 
even when there are strong correlations. On the other hand, 
if the thermal motion consists of many modes, 
the difference is larger when the correlations are stronger. 

Examples of the one-mode oscillator are given in Fig. \ref{experiment}. 
The measured thermal motion spectra of the leaf spring 
with inhomogeneous losses below 100 Hz were the same as the 
estimated values of the "traditional" modal expansion. 
This is because these fluctuations 
were dominated by only the first mode (about 60 Hz). 
As another example, 
let us consider a single-stage suspension for a mirror 
in an interferometric gravitational-wave detector. 
The sensitivity of the interferometer is limited by the thermal noise of 
the suspensions between 10 Hz and 100 Hz. Since, 
in this frequency region, this thermal noise 
is dominated by only the pendulum mode \cite{Saulson}, 
the thermal noise generated by the inhomogeneous loss 
agrees with the traditional modal expansion. 
It must be noticed that the above discussion is valid only when the 
other suspension modes are negligible. For example, 
when the laser beam spot on the mirror surface is shifted, 
the two modes (pendulum mode and mirror rotation mode) 
must be taken into account. 
In such cases, the inhomogeneous loss causes 
a disagreement between the 
real thermal noise of the single-stage suspension 
and the traditional modal expansion \cite{Braginsky}. 

The discrepancy 
between the actual thermal motion and the traditional modal expansion 
in the elastic modes 
of the mirror \cite{Yamamoto2} is larger than that of the drum, the prototype 
of the real mirror in our previous experiment \cite{Yamamoto3}. 
One of the reasons is that
the thermal motion of the 
mirror (rigid cylinder) consists of many modes \cite{Gillespie,Bondu}. 
The drum (hollow cylinder) had only two modes \cite{Yamamoto3}. 
Since the number of modes that contribute to
the thermal noise of the mirror in the interferometer 
increases when the laser beam radius becomes smaller \cite{Gillespie,Bondu}, 
the discrepancy is larger with a narrower beam. 
This consideration is consistent 
with our previous calculation \cite{Yamamoto2}. 

\subsection{Absolute value of the correlation}

\begin{figure}
\includegraphics[width=8.6cm]{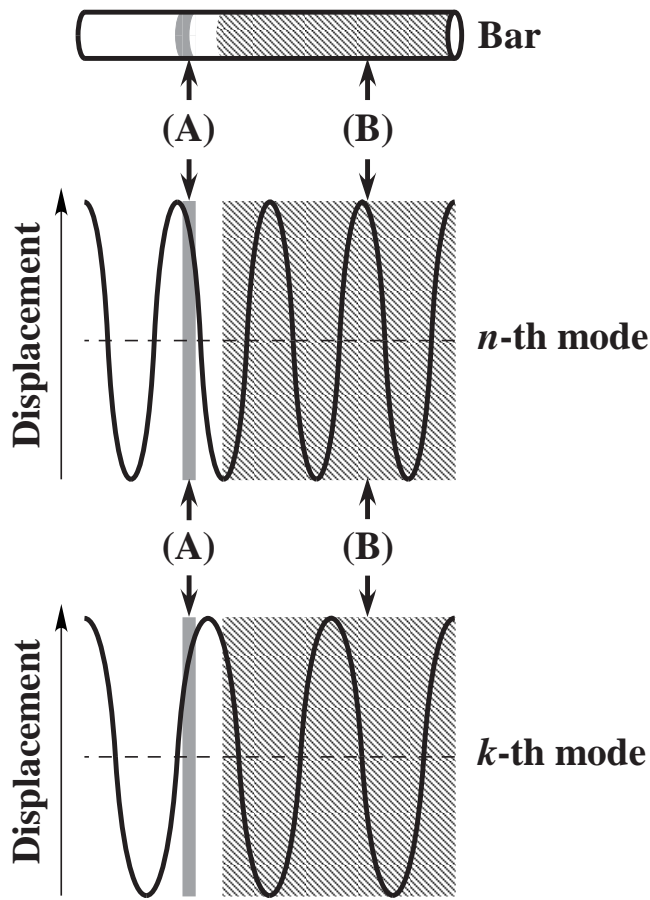}
\caption{\label{abs ex}Example for considering 
the absolute value of the coupling. 
There are the $n$-th and $k$-th modes, 
$\boldsymbol{w}_n$ and $\boldsymbol{w}_k$, 
of a bar with both free ends. 
The vertical axis is the displacement. 
The dashed horizontal lines show the bar that does not vibrate. 
When only the grey part (A), which is narrower than the wavelengths 
on the left-hand side, has viscous damping, the absolute 
value of the coupling, Eq. (\ref{coupling_viscous}), is large. 
Because the signs of $\boldsymbol{w}_n$ and $\boldsymbol{w}_k$ 
do not change in this region. 
If viscous damping exits only in the hatching part (B), which is wider 
than the wavelengths 
on the right-hand side, the coupling is about zero, 
because, in this wide region, the sign of the integrated function in 
Eq. (\ref{coupling_viscous}), which is proportional to 
the product of $\boldsymbol{w}_n$ and $\boldsymbol{w}_k$, changes.}
\end{figure}
In Eq. (\ref{G_q_n_q_k}), the absolute value of the cross-spectrum density, 
$G_{q_n q_k}$, is proportional to that 
of the coupling, $\alpha_{nk}$. 
Equations (\ref{coupling_viscous}) and (\ref{coupling_structure}) show that 
the coupling depends 
on the scale of the dissipation distribution. 
A simple example of viscous damping is shown in Fig. \ref{abs ex}. 
Let us consider the absolute value of 
$\alpha_{nk}$ 
when the viscous damping is concentrated 
(at around $\boldsymbol{r}_{\rm vis}$)
in a smaller volume ($\Delta V$) than the wavelengths 
of the $n$-th and $k$-th modes. An example of this case is (A) 
in Fig. \ref{abs ex}.
It is assumed that the vector 
$\boldsymbol{w}_n(\boldsymbol{r}_{\rm vis})$ 
is nearly parallel to $\boldsymbol{w}_k(\boldsymbol{r}_{\rm vis})$.
The absolute value of the 
coupling is derived from Eqs. (\ref{phi_n}) and (\ref{coupling_viscous}) as 
\begin{eqnarray}
|\alpha_{nk}| &\sim& |\omega \rho \Gamma(\boldsymbol{r}_{\rm vis}) 
\boldsymbol{w}_n(\boldsymbol{r}_{\rm vis}) \cdot  
\boldsymbol{w}_k(\boldsymbol{r}_{\rm vis}) \Delta V| \nonumber\\
&\sim& \sqrt{\omega \rho \Gamma(\boldsymbol{r}_{\rm vis}) 
|\boldsymbol{w}_n(\boldsymbol{r}_{\rm vis})|^2 \Delta V} \nonumber\\
&&\hspace{0.5cm}\times \sqrt{\omega \rho \Gamma(\boldsymbol{r}_{\rm vis}) 
|\boldsymbol{w}_k(\boldsymbol{r}_{\rm vis})|^2 \Delta V}\nonumber\\
&\sim& \sqrt{\alpha_{nn}\alpha_{kk}} 
= \sqrt{m_n {\omega_n}^2 \phi_n m_k {\omega_k}^2 \phi_k}.
\label{coupling_narrow}
\end{eqnarray}
The absolute value of the cross-spectrum is derived 
from Eqs. (\ref{G_q_n}), (\ref{G_q_n_q_k}), and (\ref{coupling_narrow}) as
\begin{equation}
|G_{q_n q_k}| \sim \sqrt{G_{q_n}G_{q_k}}.
\label{maxcorrelation}
\end{equation}
In short, 
the correlation is almost perfect \cite{maxcoupling}.
On the other hand, if the loss is distributed more broadly 
than the wavelengths, 
the coupling, i.e. the correlation, is about zero, 
\begin{equation}
|G_{q_n q_k}| \sim 0.
\end{equation}
The dissipation in the case where the size is larger than the 
wavelengths is equivalent to the homogeneous loss. 
An example of this case is (B) in Fig. \ref{abs ex}. 
Although the above discussion is for the case of viscous damping, 
the conclusion is also valid for other kinds of dissipation.
When the loss is localized in a small region, 
the correlations among many modes are strong. 
The loss in a narrower volume causes a 
larger discrepancy between the actual thermal motion and 
the traditional modal expansion. This conclusion coincides with 
our previous calculation of 
a mirror with inhomogeneous loss \cite{Yamamoto2}.

\subsection{Sign of correlation}

The sign of the correlation 
depends on the frequency, the loss distribution, and the position of 
the observation area. 
The position dependence provides a solution to the inverse problem: 
an evaluation of the distribution and frequency dependence 
of the loss from measurements of the 
thermal motion.

\subsubsection{Frequency dependence}

According to Eq. (\ref{G_q_n_q_k}), 
the sign of the correlation reverses at the resonant frequencies. 
For example, in calculating the double pendulum \cite{Majorana}, 
experiments involving the drum \cite{Yamamoto3} and a resonant 
gravitational wave detector with optomechanical readout \cite{Conti}, 
this change of the sign was found. 
In some cases, the thermal-fluctuation spectrum changes drastically 
around the resonant frequencies. 
A careful evaluation is necessary when the 
observation band includes the resonant frequencies. 
Examples are when using 
wide-band resonant gravitational-wave detectors 
\cite{wide1,wide2,wide3,wide4}, 
and thermal-noise interferometers \cite{Numata3,Black}. 
The reason for the reverse at the resonance is that 
the sign of the transfer function of the mode with a 
small loss from the force ($F_n$) to the 
motion ($q_n$), $H_n$ in 
Eq. (\ref{traditional3}) [$\propto (-\omega^2+{\omega_n}^2)^{-1}$], 
below the resonance is opposite to that above it.

Since the sign of the correlation changes at the resonant frequencies, 
the cross-spectrum densities, the second term of Eq. (\ref{advanced2}), 
make no contribution to the 
integral of the power spectrum density 
over the whole frequency region, i.e. the mean square 
of the thermal fluctuation, $\overline{X^2}$, as shown 
in Sec. \ref{thermal noise of advanced}. Therefore, 
the consideration in Sec. \ref{thermal noise of advanced} indicates 
that a reverse of the sign 
of the correlation conserves the equipartition principle, 
a fundamental principle in statistical mechanics.    

\subsubsection{Loss and observation area position dependence}

\begin{figure}
\includegraphics[width=8.6cm]{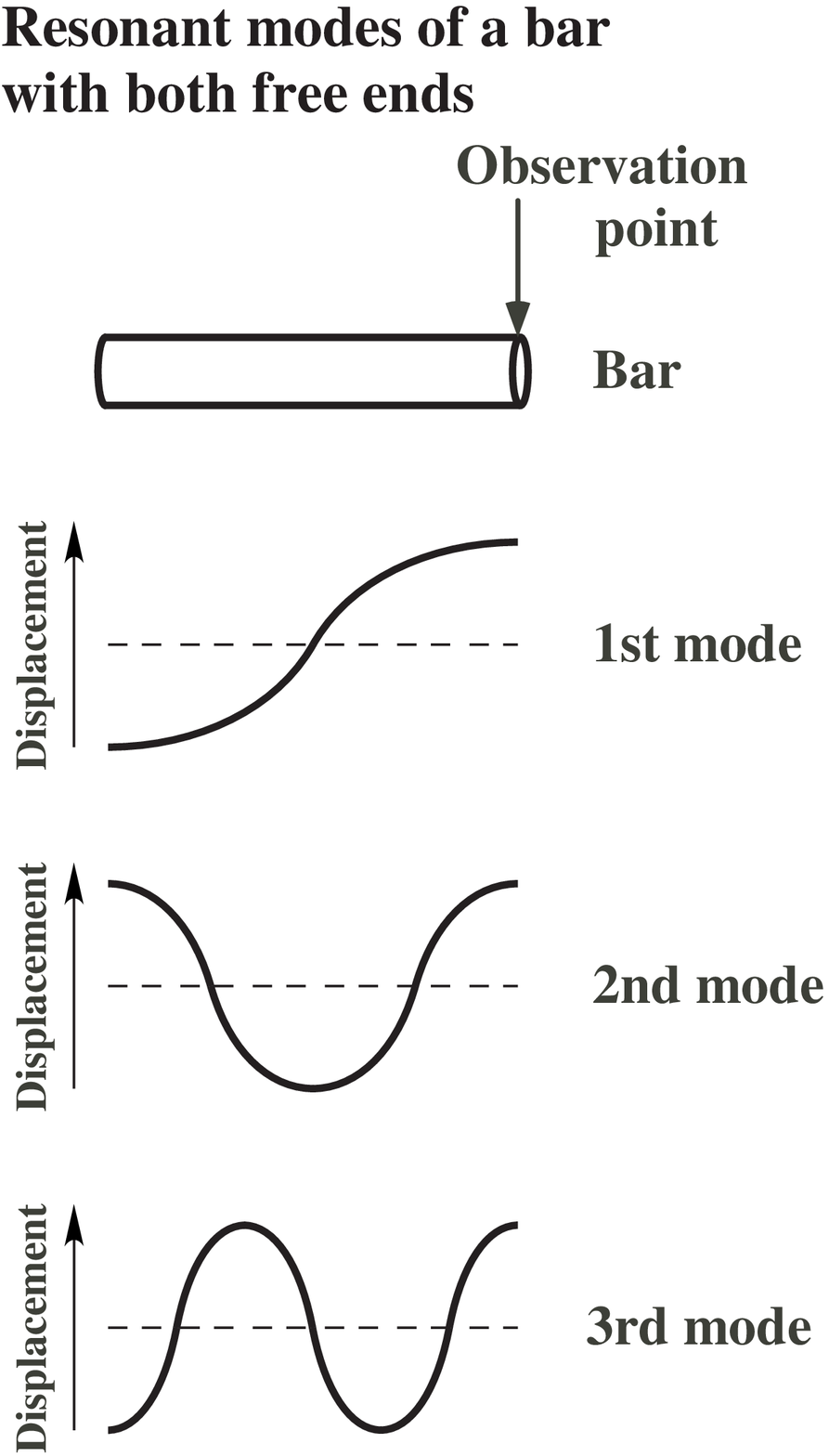}
\caption{\label{signex}Example for considering the sign of 
the coupling. There are the lowest three modes, $\boldsymbol{w}_n$, 
of a bar with both free ends. 
The vertical axis is the displacement. 
The dashed horizontal lines show the bar that does not vibrate. 
The observation point is 
at the right-hand side end.
The normalization condition is Eq. (\ref{normalized condition}) 
\cite{normalized condition}. 
The sign and shape of the displacement of 
all the modes around the observation point are positive and similar, 
respectively. 
On the contrary, at the left-hand side end, the sign and shape of the $n$-th 
mode are different from each other in many cases. 
From Eqs. (\ref{normalized condition}), 
(\ref{coupling_viscous}), (\ref{coupling_structure}),
when the loss is concentrated near to the observation area, 
most of the couplings are positive. 
On the other hand, when the loss is localized far from the 
observation area, the number of the negative couplings is 
about the same as the positive one. 
In such a case, most of the couplings 
between the $n$-th and $(n \pm 1)$-th modes are negative.}
\end{figure}
According to Eqs. (\ref{coupling_viscous}) and (\ref{coupling_structure}), 
and the normalization condition, Eq. (\ref{normalized condition}) 
\cite{normalized condition}, 
the sign of the coupling, $\alpha_{nk}$, 
depends on the loss distribution and 
the position of the observation area. 
A simple example is shown in Fig. \ref{signex}. 
Owing to this normalization condition, 
near the observation area, the basis functions, $\boldsymbol{w}_n$, 
are similar in most cases.
On the contrary, in a volume far from the observation area, 
$\boldsymbol{w}_n$ is different from each other in many cases. 
From Eqs. (\ref{normalized condition}), 
(\ref{coupling_viscous}), (\ref{coupling_structure}) and 
(\ref{G_F_n_F_k}),
when the loss is concentrated near to the observation area, 
most of the couplings (and the correlations 
between the fluctuation forces of the modes, $G_{F_nF_k}$) are positive. 
On the other hand, when the loss is localized far from the 
observation area, the numbers of the negative couplings and $G_{F_nF_k}$ are 
about the same as the positive ones. In such a case, most of the couplings 
between the $n$-th and $(n \pm 1)$-th modes (and $G_{F_nF_{n \pm 1}}$)
are negative. These are because the localized loss tends 
to apply to the fluctuation force 
on all of the modes 
to the same direction around itself.
Equation (\ref{G_q_n_q_k}) indicates that 
the sign of the correlation, $G_{q_nq_k}$, is the same 
as that of the coupling, $\alpha_{nk}$, below the first resonance. 
In this frequency band, the thermal motion is larger and smaller than the
evaluation of the traditional modal expansion if the dissipation is 
near and far from the observation area, respectively. 
This conclusion is consistent 
with the qualitative discussion of Levin \cite{Levin}, 
our previous calculation of the mirror \cite{Yamamoto2}, 
and the drum experiment \cite{Yamamoto3}.

\subsubsection{Inverse problem}

The above consideration about the sign of the coupling gives 
a clue to solving the inverse problem: 
estimations of the 
distribution and frequency dependences of the loss from the 
measurement of the thermal motion. 
Since the sign of the coupling 
depends on the position of the 
observation area and the loss distribution, 
a measurement of the thermal vibrations at multiple points 
provides information about the couplings, 
i.e. the loss distribution. 
Moreover, multiple-point measurements 
reveal the loss frequency dependence. 
Even if the loss is uniform, 
the difference between the actual thermal motion 
and the traditional modal expansion exits 
when the expected frequency dependence of the loss angles, $\phi_n(\omega)$, 
is not correct \cite{Majorana}.
The measurement at the multiple points shows 
whether the observed difference 
is due to an inhomogeneous loss or 
an invalid loss angle. This is because the sign of the difference 
is independent of the position of the observation area 
if the expected loss angles are not valid.

As an example, our leaf-spring experiment \cite{Yamamoto1} is discussed. 
The two graphs on the right (or left) side of Fig. \ref{experiment} show 
thermal fluctuations at different positions in the same mechanical system. 
The spectrum is smaller than the traditional modal expansion. 
The other one is larger. 
Thus, the disagreement in the leaf-spring experiment was due to 
inhomogeneous loss, not invalid loss angles. When the power 
spectrum had a dip between the first (60 Hz) and second (360 Hz) modes, 
the sign of the correlation, $G_{q_1q_2}$, was negative. 
According to Eq. (\ref{G_q_n_q_k}), the sign of the coupling, 
$\alpha_{12}$, was 
positive. The loss was concentrated near to the observation point 
when a spectrum dip was found. 
The above conclusion agrees 
with the actual loss shown in Fig. \ref{experiment}.

\section{Conclusion}

The traditional modal expansion has frequently been used 
to evaluate the thermal noise of 
mechanical systems \cite{Saulson}. 
However, recent experimental research 
\cite{Yamamoto1,Harry,Conti,Numata3,Yamamoto3,Black} has proved that 
this method is invalid when the mechanical dissipation 
is distributed inhomogeneously. 
In this paper, 
we introduced a modification of the modal expansion 
\cite{Yamamoto-D,Majorana}. 
According to this method 
(the advanced modal expansion), 
inhomogeneous loss causes correlations between the thermal 
fluctuations of the modes. 
The fault of the traditional modal expansion is that 
these correlations are not taken into account. 
Our previous experiments \cite{Yamamoto1, Yamamoto3} concerning 
the thermal noise of the inhomogeneous loss support the 
advanced modal expansion. 

The advanced modal expansion gives interesting physical insight 
about the difference between 
the actual thermal noise and the traditional modal expansion. 
When the thermal noise 
consists of the contributions of many modes, the loss is localized in 
a narrower area, which makes a larger difference.
When the thermal noise is dominated by only one mode, this difference is 
small, even if the loss is extremely inhomogeneous. 
The sign of this difference 
depends on the 
frequency, the distribution of the loss, 
and the position of the observation area. 
It is possible to derive the distribution and frequency dependence 
of the loss from measurements of 
the thermal vibrations at multiple points. 

There were many problems concerning 
the thermal noise caused by inhomogeneous loss.
Our previous work \cite{Yamamoto1,Yamamoto3,Yamamoto2} 
and this research solved almost all of these problems: 
a modification of the traditional estimation method (in this paper), 
experimental checks of the new and traditional estimation methods 
and a confirmation of the new methods 
(\cite{Yamamoto1,Yamamoto3} and this paper), 
an evaluation of the thermal noise of the 
gravitational wave detector using the new method \cite{Yamamoto2}, 
and a consideration of the physical properties of the 
discrepancy between the actual thermal noise and 
the traditional estimation method 
(in this paper).

\begin{acknowledgments}
This research was supported in part by Research Fellowships of the Japan
Society for the Promotion of Science for Young Scientists, and 
by a Grant-in-Aid for Creative Basic Research of the Ministry of
Education. 
\end{acknowledgments}


\begin{thebibliography}{99}
\bibitem{alpha2}In the off-resonance region, where 
$|-\omega^2+{\omega_n}^2| \gg {\omega_n}^2 \phi_n(\omega)$ for all $n$,  
this approximation is always appropriate 
because the maximum of ${\alpha_{nk}}^2$ is 
$m_n {\omega_n}^2 \phi_n m_k {\omega_k}^2 \phi_k$ \cite{maxcoupling}. 
In the calculation to derive Eq. (\ref{advanced3}), 
Cramer's rule is useful.
\bibitem{KKcomment}The sign on the right-hand side 
of the Kramers-Kronig relation 
in Ref. \cite{Landau2} is positive. 
The definition of the Fourier transformation 
in Ref. \cite{Landau2} is conjugate to that of this paper, 
Eq. (\ref{Fourier transform}).
\bibitem{maxcoupling}This discussion 
suggests that $|\alpha_{nk}|$ is never larger than 
$\sqrt{m_n {\omega_n}^2 \phi_n m_k {\omega_k}^2 \phi_k}$. 
The validity of this conjecture is supported 
by the Cauchy-Schwarz inequality \cite{Yamamoto-D}.
\bibitem{normalized condition}
The results of the 
discussion presented in this section are valid under 
arbitrary normalization conditions. 
If a normalization condition other 
than Eq. (\ref{normalized condition}) is adopted, 
the signs of some couplings change. 
The signs of the right-hand sides of 
Eqs. (\ref{observed coordinate decomposition}) 
and (\ref{advanced1}) also change. 
Equations (\ref{observed coordinate}), (\ref{eq_mo_continuous}), 
and (\ref{mode decomposition}) say that 
the right-hand sides of 
Eqs. (\ref{observed coordinate decomposition}) 
and (\ref{advanced1}) in the general style are 
$\sum_{n} q_n(t) \int \boldsymbol{w}_n(\boldsymbol{r}) 
\cdot \boldsymbol{P}(\boldsymbol{r}) dS$
and $\tilde{F} \int \boldsymbol{w}_n(\boldsymbol{r}) 
\cdot \boldsymbol{P}(\boldsymbol{r}) dS$, respectively.
The change in the sign cancels each other in the process of the 
calculation for the transfer function, $H_X$, and the power spectrum of 
the thermal motion, $G_{X}$.
\end{thebibliography}

\end{document}